\documentclass[conference]{IEEEtran}
\usepackage{cite}
\usepackage{amsmath,amssymb,amsfonts}
\usepackage{balance}
\usepackage{algorithmic}
\usepackage{graphicx}
\usepackage{caption}
\usepackage{subcaption}
\usepackage{textcomp}
\usepackage{xcolor}
\usepackage{tabularx}
\usepackage{booktabs}
\usepackage{multirow}
\usepackage{collcell}
\usepackage{hhline}
\usepackage[normalem]{ulem}
\newcommand{\ru}{\rule{0mm}{3mm}}
\def\BibTeX{{\rm B\kern-.05em{\sc i\kern-.025em b}\kern-.08em
    T\kern-.1667em\lower.7ex\hbox{E}\kern-.125emX}}
\begin{document}

\title{Deepfake audio detection by speaker verification }

\author{
\IEEEauthorblockN{Alessandro Pianese, Davide Cozzolino, Giovanni Poggi and Luisa Verdoliva} 
\IEEEauthorblockA{University Federico II of Naples, Italy \\
Email: \{name.surname\}@unina.it}
}
\maketitle

\begin{abstract}
Thanks to recent advances in deep learning, sophisticated generation tools exist, nowadays, that produce extremely realistic synthetic speech. However, malicious uses of such tools are possible and likely, posing a serious threat to our society. Hence, synthetic voice detection has become a pressing research topic, and a large variety of detection methods have been recently proposed. Unfortunately, they hardly generalize to synthetic audios generated by tools never seen in the training phase, which makes them unfit to face real-world scenarios. 
In this work we aim at overcoming this issue by proposing a new detection approach that leverages only the biometric characteristics of the speaker, with no reference to specific manipulations. Since the detector is trained only on real data, generalization is automatically ensured. The proposed approach can be implemented based on off-the-shelf speaker verification tools. We test several such solutions on three popular test sets, obtaining good performance, high generalization ability and high robustness to audio impairment.  
\end{abstract}


\section{Introduction}

Recent methods for synthetic voice generation are becoming more and more sophisticated.
State-of-the-art methods for Text to Speech (TTS) synthesis and Voice Conversion (VC) allow one to generate extremely realistic audio waveforms with such a high level of naturalness to fool both humans and automatic speaker verification (ASV) systems \cite{Das2020attacker}.
Convincing fake audios, when used for malicious purposes, pose a serious threat to our society, as they allow us to impersonate people or create deepfake videos simply by synchronizing the generated speech with the movement of the lips.

In the current literature, several detection methods have been proposed, most of them based on supervised learning. 
Traditional methods typically rely on the extraction of acoustic features in the spectral domain.
In fact, through time-frequency analysis it is possible to obtain a rich and expressive representation of the voice, based on the well-known frequency cepstral coefficients or constant Q cepstral coefficients \cite{wang2021comparative,Zhang2021fake}. 
Also effective are first- and second-order spectral features \cite{Sahidullah2015comparison} 
as well as higher-order polyspectral features \cite{AlBadawy2019detecting}, 
exploiting the observation that synthesized speech can introduce uncommon spectral correlations. 
Other approaches use features related to auto-regressive modeling of the speech \cite{Janicki2015spoofing} 
or a combination of spatial and spectral features \cite{Tak2021end, Borrelli2021synthetic},
or rely on a constrained convolutional neural network
where the first layer act as a bank of band-pass filters 
with a fixed rectangular-shaped filter response and learned cut-in and cut-off frequencies \cite{Tak2021rawnet2}.  

These approaches are usually very effective on the data they are trained for, but suffer a significant, even dramatic, performance impairment when used on data synthesized by generation models never seen in training.
This is a major weakness in  real-world scenarios, where the manipulation attack is not known in advance. 
Therefore, it is extremely important to devise methods that keep working well also "in the wild".
Indeed, one major goal of the 2019 ASVspoof initiative  
was to foster spoofing countermeasures that generalize smoothly to unseen attacks. 
To this end, a dataset was created where 
the training and development partitions included a set of 6 attacks, 
while the evaluation partition included a different set of 13 attacks only partially overlapping with the former. 

Given the fast progress of text-to-speech technologies,
it is important to understand to what extent methods trained for specific attacks generalize to new ones,
and try to ensure a good behavior for them also in real-world scenarios.
A possible way to improve generalization is to consider a large number of attacks in training or,
equivalently, to fuse multiple detection systems, each tuned to a different attack.
However, this amounts only to moving the boundary a little further, without really solving the problem. The system will keep being prone to new unseen attacks.

\begin{figure*}[t!]
    \centering
    \includegraphics[width=0.95\linewidth, trim=0 150 0 0, clip, page=1]{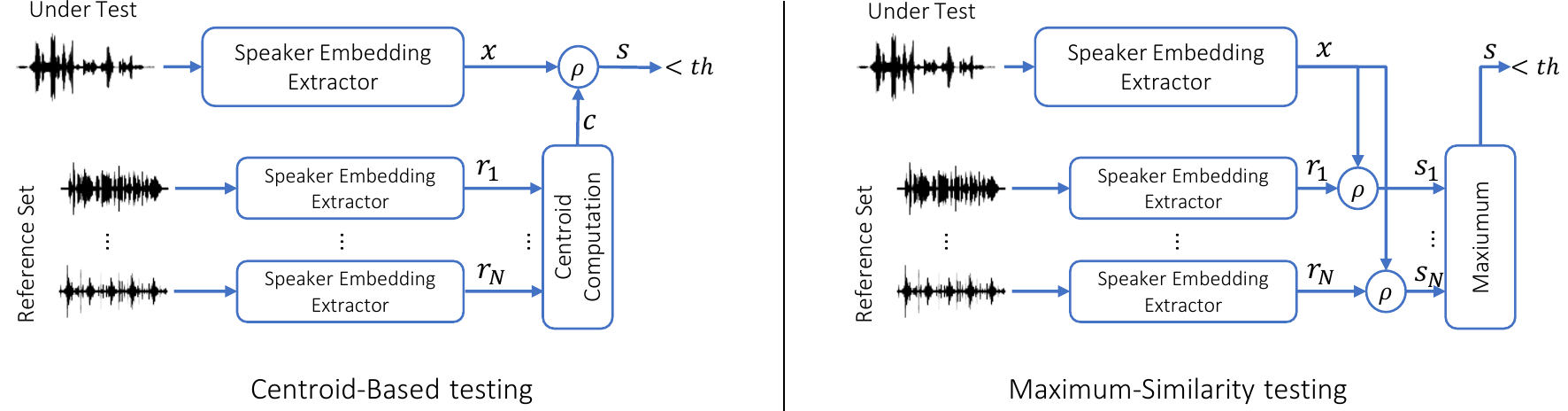}
    \caption{Schemes of the two testing strategies: Centroid-Based (CB) and  Maximum-similarity (MS). In both strategies, we project the audio under-test and all the audios of the reference-set in the embedding space by the speaker embedding extractor.  
    On the left we show the CB testing strategy: a centroid vector ($c$) is computed averaging all the embedded vectors of the reference-set  ($r_i$) and the embedded vector of the audio under-test ($x$) is compared through the similarity metric ($\rho$) only with the centroid one. The similarity ($s$) obtained in this way is used as decision statistic to detect deepfake audios.
    On the right we show the MS testing strategy: the embedded vector of the audio under-test ($x$) is compared with each embedded vector of the reference-set ($r_i$) through the similarity metric ($\rho$) and the maximum similarity ($s$) is provided as decision statistic.
    }
    \label{fig:testing}
\end{figure*}

A more radical solution to the generalization issue is to rely on one-class methods \cite{Zhang2021one}.
By definition, these are trained only on the class of interest, the real class in our case, and data that depart from it are considered as anomalies, that is, fake.
Since no fake data are used in training, there is no dependence on any specific form of manipulation, and generalization is ensured by default.
Of course, a description of the class of real audios so precise to enable reliable anomaly detection is out of reach.
However, the idea of training only on real data can be implemented in other forms.
For example, 
several papers have recently appeared \cite{Agarwal2019, Agarwal2020, Cozzolino2021idreveal, Cozzolino2022audiovisual} where video deepfakes are detected by means of a person-of-interest approach.
The detector tries to establish if the subject of the video is really the person that is claimed to be: if not, the video is labeled as fake.
Therefore, rather than looking for traces of manipulation, these methods leverage high-level biometric features.
Since training takes place only on original data, generalization is guaranteed.
In \cite{Agarwal2019, Agarwal2020, Cozzolino2021idreveal} only video biometric features are exploited, while \cite{Cozzolino2022audiovisual} exploits both audio and video features.

Here, we investigate the potential of the person-of-interest (PoI) approach when only the audio track is available or exploitable.
We assume to know the identity of the speaker of the audio under analysis and to have a reference set of pristine audio tracks for such identity.
Then, the identity is verified by comparing suitably extracted embedded vectors of test and reference audios.
We note that this differs from \cite{Castan2022speaker}, where the proposed supervised approach is fine-tuned on real and synthetic speech of the identity under test.
Instead, it can be observed that, with our approach, synthetic audio detection boils down to a speaker verification problem.
This is true, and in fact we adapt several state-of-the-art speaker verification methods to perform fake audio detection and study their performance in several conditions of interest. 

In summary, our main contributions are:
\begin{itemize}
\item we show that person-of-interest methods, including speaker verification approaches, can be used to effectively detect synthesized speech, even in the presence of noise;
\item we prove this solution to largely outperform state-of-the-art supervised methods in terms of generalization ability.
\end{itemize}

\begin{table*}[!t]
    \centering
    \begin{tabular}{lccccccc}
    \toprule
\ru ~~~~~~~Method                                           & Input & Architecture   & Pooling & Aug. & Loss function & Feature len. & Similarity metric \\ \midrule
\ru \cite{Chung2020in} ClovaAI                          & 40-MelSpec & Thin ResNet-34           & SAP &  No & AP & 512 &  Cosine Similarity \\
\ru \cite{Heo2020clova} H/ASP                           & 64-MelSpec & Half ResNet-34           & ASP & Yes & AP+softmax & 512 &  Cosine Similarity \\
\ru \cite{desplanques2020ecapa} ECAPA-TDNN              & 80-MelSpec & SqueezeExcitation ResNet & ASP & Yes & AAM-softmax & 192 &  Cosine Similarity \\
\ru \cite{Cozzolino2022audiovisual} POI-Forensics       & 257-Spec   & GN ResNet-50             & GAP &  No & Contr. loss & 256 & Squared Euclidean \\
\ru \cite{Cozzolino2022audiovisual} POI-Forensics + aug & 257-Spec   & GN ResNet-50             & GAP & Yes & Contr. loss & 256 & Squared Euclidean \\
\bottomrule
    \end{tabular}
    \vspace{2mm}
    \caption{Characteristics of the speaker verification techniques used in this work.}
    \label{tab:methods}
\end{table*}

\section{Synthetic audio detection workflow}

We want to evaluate the potential of audio speaker verification techniques for deepfake audio detection.
In our scenario, the {\em claimed} identity of the speaker is known, and a reference set, $\mathcal{R}$, of real audios of the same identity is available.
Our procedure is to compare the audio under test with those of the reference set in the embedding space of the audio speaker verification.
To this aim, we consider two different testing strategies, shown in figure \ref{fig:testing}: Centroid-Based (CB) and Maximum-Similarity (MS).
The first strategy is inspired by Centroid-Based Metric Learning \cite{Wang2019centroid}.
We first compute the centroid $c = \sum_{i\in \mathcal{R}} r_i / |\mathcal{R}|$ of the reference set embedding vectors $\left\{r_i \right\}_{i\in \mathcal{R}}$.
Then, the similarity, $s=\rho (x,c)$, between this centroid and the embedding vector of the audio under test, $x$, is used as decision statistic, and the audio is labeled as fake if $s$ is below a given threshold.
In all cases, $\rho (\cdot,\cdot)$ is the similarity metric adopted by the speaker verification technique. 
In the Maximum-Similarity testing strategy, proposed in \cite{Cozzolino2021idreveal, Cozzolino2022audiovisual}, a similarity index, $s_i=\rho (x,r_i)$, is evaluated between the audio under test and each audio of reference set. 
Then, only the maximum $s=\max_{i\in \mathcal{R}} s_i$ is considered as a decision statistic.

We consider various audio speaker verification techniques, 
they are described below and a summary of their characteristics is reported in table \ref{tab:methods}.

The \textbf{Clova-AI} \cite{Chung2020in} method is named after the proposing group, involved in the development of the Clova personal assistant for Android.
The main focus is on metric learning, considered especially well-fit to deal with a realistic open-set scenario.
In input, the model accepts a 40-dimensional Mel Spectrogram with a 25ms window width and a 10 ms frame shift.
The backbone is Thin ResNet-34 which is the same as ResNet-34\cite{he2016deep} except for using only one-quarter of the channels in each residual block.
Further minor modifications lead to Fast ResNet-34, with reduced complexity.
Self-Attentive Pooling (SAP) \cite{cai2018exploring} is used to aggregate frame-level features into utterance-level representation, with larger weights associated with more informative frames.
The authors introduce also a new loss function, called Angular Prototypical (AP) loss,
that chooses some elements of the batch to obtain centroids for each speaker, and uses cosine distance to obtain a similarity score for each sample.

In \cite{Heo2020clova}, researchers of the same group report on their proposals for the 2020 VoxCeleb Speaker Recognition Challenge (VoxSRC).
The best experimental results are obtained with the \textbf{H/ASP} method with AP+Softmax loss.
64-dimensional log Mel-filterbanks are now used as the input to the network.
The backbone is again a variant of ResNet-34 with one-half (as opposed to one-quarter) channels in each residual block. 
Attentive Statistic Pooling (ASP) \cite{okabe2018attentive} is used, where the channel-wise weighted standard deviation concurs with the weighted mean for temporal-frame aggregation.
The AP loss proposed in \cite{Chung2020in} is combined with a standard softmax loss.
The method uses two types of augmentation: additive background noise and reverberation simulation.
In the first case, background audios (ambient sounds, human speech, and music) from the MUSAN dataset \cite{Snyder2015musan} are used, with a random signal-to-noise-ratio (SNR) going from 5 to 20 dB.
For reverberation simulation, the simulated room impulse response filters provided in \cite{Ko2017study} are used.

The \textbf{ECAPA-TDNN} \cite{desplanques2020ecapa} method is based on a Time Delay Neural Network (TDNN), with several improvements with respect to previous solutions.
First, Squeeze-and-Excitation (SE) blocks are introduced to explicitly model channel interdependencies. 
Together with convolutional, block normalization and ReLU layers, these are included in the core SE-Res2Blocks residual blocks.
Then, ASP is employed to select the most insightful features, and the Angular Margin Softmax (AAM-softmax) \cite{Deng2019arcface} is used as loss function.
As input, the model accepts 80-dimensional Mel Spectrogram, again with a 25 ms window width and a 10 ms frame shift.
Augmentation strategies similar to those of H/ASP are used.

\textbf{POI-Forensics} was proposed in \cite{Cozzolino2022audiovisual} to perform deepfake detection based on audio-visual information with a person-of-interest (PoI) approach.
A deepfake is detected when the person in the video turns out not to be who is claimed to be.
In other words, deepfake detection is explicitly recast as a person identification problem, based on soft biometrics information, with decision relying jointly on video and audio clues.
Here, we neglect the video information and use exclusively the audio track, performing therefore speaker verification.
In \cite{Cozzolino2022audiovisual}, the audio feature extraction network is based on a ResNet-50 backbone with Group-Normalization and Global Average Pooling.
A crucial role is played by the contrastive learning formulation, aimed at embedding vectors of the test audio close to other embedded vectors of the same subject, but far from those of different subjects.
In particular, a Distance-Based Logistic Loss is adopted, proposed in \cite{Khosla2020supervised} and widely used in the literature.
As input, the network accepts a 257-dimensional Spectrogram with a 25 ms window width and a 10 ms frame shift.
Here, we consider also a variant of the method with augmentation, obtained by adding background noise with random SNR going from 3 to 25 dB.

\section{Experimental results}

In this Section, 
we describe the supervised methods for audio spoofing detection used as state-of-the-art references,
the datasets used in the experiments,
and the experimental results, with special emphasis on generalization and robustness.

\subsection{Reference methods} 

\textbf{RawNet2-antispoofing} \cite{Tak2021rawnet2} processes the waveform by a bank of band-pass filters. Afterwards, this model makes use of residual blocks but with two key differences: firstly, each residual block contains a filter-wise feature map scaling to enhance the discrimination ability of the features, as shown in \cite{woo2018cbam}; secondly, before the fully connected layer, a gated recurrent unit is used to aggregate features to form a single utterance-level representation.

\textbf{RawGAT-ST-mul}\cite{Tak2021end} is a model that aims to exploit artifacts in specific subbands and temporal portions of a recording. Also this model uses a bank of band-pass filters, whose output is first fed to several residual blocks and then is given as input to the separable spectral-temporal attention graph layer. The resulting features are merged by element-wise multiplication before being fed to the unified spectral-temporal attention layer.

\textbf{LCNN-LSTM-sum-p2s}\cite{wang2021comparative} is based on the light convolutional neural network proposed in \cite{lavrentyeva2019stc}. It has been improved by adding two bidirectional LSTM layer with skip connections. This method accepts as input 60-dimensional Linear Frequency Cepstral Coefficients (LFCCs) with a frame window of 20 ms and a frame shift of 10 ms. 

\subsection{Datasets}

To evaluate the methods described above, we selected three public datasets that provide information about the subject.

\textbf{ASVSpoof2019}\cite{Wang2020asv} was released as part of the 2019 Automatic Speaker Verification Spoofing And Countermeasures Challenge.
The dataset is divided into different parts, and we consider only the evaluation set formed by audio tracks manipulated with 19 different speech synthesis and voice conversion methods.
The selected part contains 7,355 pristine and 63,882 manipulated utterances from 67 different subjects. 
To build the reference sets, we adopt a leave-one-out strategy, that is, all the real utterances of the POI are included in the reference set except the audio under test.

\textbf{FakeAVCelebV2}\cite{Khalid2021fakeavceleb} is an audio and video deepfake dataset created from the videos of the VoxCeleb2 dataset \cite{Chung2018voxceleb2}. Videos are divided into four self-explanatory categories: RealVideo-RealAudio, FakeVideo-RealAudio, RealVideo-FakeAudio, and FakeVideo-FakeAudio. Fake speech tracks are generated using the real-time voice cloning tool SV2TTS\cite{Jia2018transfer}. We extracted audio tracks only from the RealVideo-RealAudio and RealVideo-FakeAudio categories, since the same audio tracks are associated with fake videos. Each category contains 500 files, so we obtained 1000 audio clips. For the reference set, we use 10 pristine audios of the subject of interest drawn from VoxCeleb2.

The \textbf{In-The-Wild Audio Deepfake dataset (IWA)}\cite{Muller2022does} contains 20.8 hours of pristine recordings and 17.2 hours of fake audios, for a total of 31,779 files, belonging to 54 English-speaking identities. Fake audios were not generated by the authors but collected from publicly available material. To ensure consistency, the authors collected real material of the same type as the fake one, e.g., a fake speech from Richard Nixon prompted them to gather a real speech from the same politician. Overall, the dataset contains 19,963 pristine utterances and 11,816 manipulated ones of 54 different subjects. Also for this dataset we adopt the leave-one-out strategy to build the reference sets. 

\subsection{Experimental analysis} 

In Table \ref{tab:results}, we report the results on the three datasets in terms of Equal Error Rate (EER), Area Under the ROC Curve (AUC), 
and minimum normalized tandem Detection Cost Function (t-DCF) \cite{Kinnunen2018tdcf}. 
The t-DCF is the weighted sum of the missing detection rate and the false alarm rate. 
We use the weights provided in the ASVspoof 2019 challenge \cite{Yamagishi2019asv} and report the minimum  t-DCT value over the decision threshold.
Note that all the speaker verification techniques are trained on the VoxCeleb2 dataset \cite{Chung2018voxceleb2}, 
while the supervised methods are trained on the ASVSpoof2019 \cite{Wang2020asv} training dataset. 
For all the techniques, we analyzed the first 4 seconds of the audio signal.

On the ASVSpoof2019 dataset, the supervised baselines obtain a near perfect performance,
which is not surprising, as they were developed for this dataset.
However, the picture changes completely when other, off-training, datasets are considered.
Only LCNN-LSTM-sum-p2 keeps providing an acceptable performance, while the other two methods degrade to the coin-tossing level.  
On the contrary, the proposed framework proves to work almost equally well across all the datasets, 
with a generally good and sometimes excellent performance, 
which degrades somewhat only on the IWA dataset, due to its intrinsically more challenging nature.
In this latter case, the Maximum-similarity testing strategy seems to be preferable,
while the Centroid-Based testing strategy ensures a better performance on ``simpler'' datasets.
This can be explained by noting that audios collected in the wild (from social networks or video streaming platforms) are likely to have a more scattered distribution in the embedding space and evaluating the similarity for each reference audio may increase the probability of a good matching.
POI-Forensics provides the best overall performance, but the speaker verification methods are not far, especially H/ASP, which is a remarkable result, considering that they were adapted on the fly to perform deepfake audio detection.

\setlength{\tabcolsep}{4.5pt}
\begin{table*}[]
    \centering
    \begin{tabular}{clcccccccccccc}
    \toprule
\ru &              & \multicolumn{3}{c}{ASVSpoof2019} & \multicolumn{3}{c}{FakeAVCeleb} & \multicolumn{3}{c}{IWA} & \multicolumn{3}{c}{average} \\
\cmidrule(lr){3-5} \cmidrule(lr){6-8} \cmidrule(lr){9-11} \cmidrule(lr){12-14}
\ru &   Method     & EER    & t-DCF & AUC    & EER  & t-DCF  & AUC & EER    & t-DCF & AUC & EER   & t-DCF & AUC \\
    \midrule
\ru \multirow{3}{2cm}{\centering Supervised methods}
    &        \cite{wang2021comparative} LCNN-LSTM-sum-p2s & 0.03 & 0.08 & 99.4 & 0.39 & 0.84 & 67.3 & 0.33 & 0.99 & 72.1 & 0.25 & 0.64 & 79.6 \\
\ru &                       \cite{Tak2021end} RawGATstMul & \textbf{0.01} & \textbf{0.03} & \textbf{99.9} & 0.66 & 1.00 & 30.5 & 0.53 & 0.99 & 47.4 & 0.40 & 0.68 & 59.2 \\
\ru &                       \cite{Tak2021rawnet2} RawNet2 & 0.21 & 0.70 & 86.1 & 0.55 & 0.99 & 45.5 & 0.51 & 1.00 & 50.7 & 0.42 & 0.89 & 60.8 \\
\midrule
\ru \multirow{5}{2cm}{\centering Centroid-Based testing}
    &                           \cite{Chung2020in} ClovaAI & 0.22 & 0.65 & 85.8 & 0.42 & 1.00 & 60.1 & 0.36 & 0.93 & 67.8 & 0.33 & 0.86 & 71.2 \\
\ru &                            \cite{Heo2020clova} H/ASP & 0.17 & 0.54 & 90.5 & \textbf{0.11} & \textbf{0.32} & \textbf{94.5} & 0.27 & 0.71 & 78.5 & 0.19 & 0.52 & 87.8 \\
\ru &               \cite{desplanques2020ecapa} ECAPA-TDNN & 0.20 & 0.62 & 87.6 & 0.17 & 0.46 & 90.4 & 0.30 & 0.79 & 75.4 & 0.22 & 0.62 & 84.4 \\
\ru &        \cite{Cozzolino2022audiovisual} POI-Forensics & 0.07 & 0.21 & 97.6 & 0.22 & 0.61 & 86.1 & 0.25 & 0.61 & 81.4 & 0.18 & 0.48 & 88.4 \\
\ru & \cite{Cozzolino2022audiovisual} POI-Forensics + aug. & 0.14 & 0.40 & 93.2 & 0.25 & 0.67 & 83.1 & 0.27 & 0.69 & 78.6 & 0.22 & 0.59 & 85.0 \\
\midrule
\ru \multirow{5}{2cm}{\centering Maximum-Similarity testing}
    &                        \cite{Chung2020in} ClovaAI    & 0.24 & 0.74 & 84.2 & 0.40 & 0.98 & 62.8 & 0.31 & 0.93 & 74.1 & 0.32 & 0.89 & 73.7 \\
\ru &                         \cite{Heo2020clova} H/ASP    & 0.20 & 0.65 & 88.3 & 0.14 & 0.41 & 93.5 & 0.22 & 0.63 & 85.2 & 0.19 & 0.56 & 89.0 \\
\ru &               \cite{desplanques2020ecapa} ECAPA-TDNN & 0.24 & 0.74 & 84.8 & 0.21 & 0.61 & 86.2 & 0.27 & 0.81 & 79.2 & 0.24 & 0.72 & 83.4 \\
\ru &        \cite{Cozzolino2022audiovisual} POI-Forensics & 0.10 & 0.32 & 96.1 & 0.21 & 0.58 & 85.8 & \textbf{0.15} & \textbf{0.39} & \textbf{91.9} & \textbf{0.15} & \textbf{0.43} & \textbf{91.3} \\
\ru & \cite{Cozzolino2022audiovisual} POI-Forensics + aug. & 0.18 & 0.54 & 90.4 & 0.26 & 0.65 & 82.4 & 0.19 & 0.53 & 88.2 & 0.21 & 0.58 & 87.0 \\
\bottomrule
    \end{tabular}
    \vspace{2mm}
    \caption{Results of all considered methods on all datasets in term of EER, minimum t-DCF and AUC. The supervised methods are trained on the development set of ASVSpoof2019\cite{Wang2020asv}, while the speaker verification method are trained on the VoxCeleb2 dataset \cite{Chung2018voxceleb2}. }
    \label{tab:results}
\end{table*}

\begin{figure}
    \centering
    \includegraphics[width=0.9\linewidth]{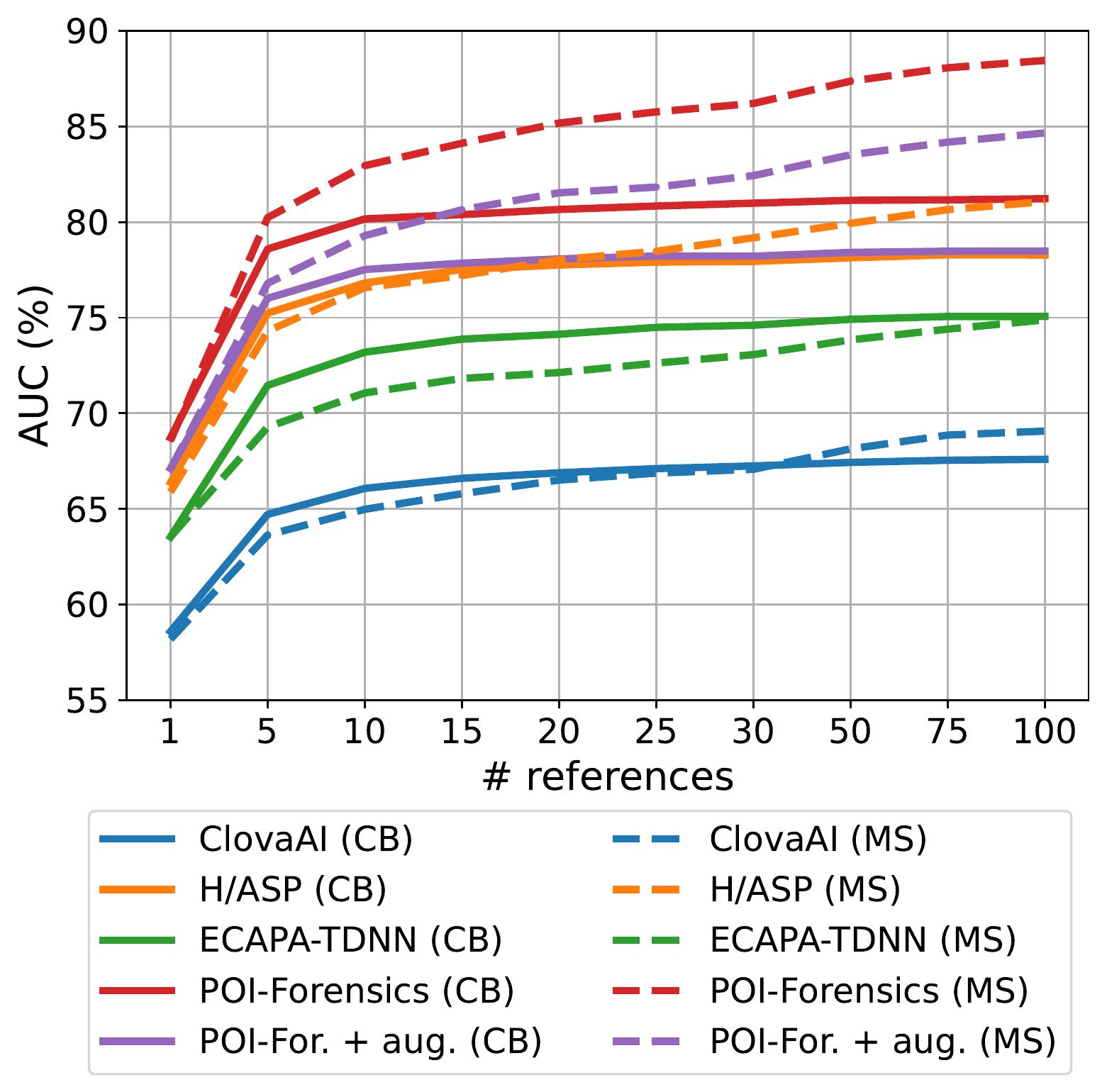}
    \caption{AUC performance on the IWA dataset as a function of the number of elements in the Reference Set. With the Maximum-Similarity testing strategy, the performance keeps growing steadily with no sign of saturation.}
    \label{fig:ref}
\end{figure}

In figure \ref{fig:ref}, only for the IWA dataset, we show the AUC performance as a function of the number of elements of the reference set. As expected, the performance improves with increasing cardinality. However, with Centroid-Based testing strategy the performance plateaus at about 20 elements, while it keeps growing steadily with the Maximum-Similarity testing strategy, confirming the above observations.

\begin{figure*}
    \centering
    \includegraphics[width=0.85\linewidth]{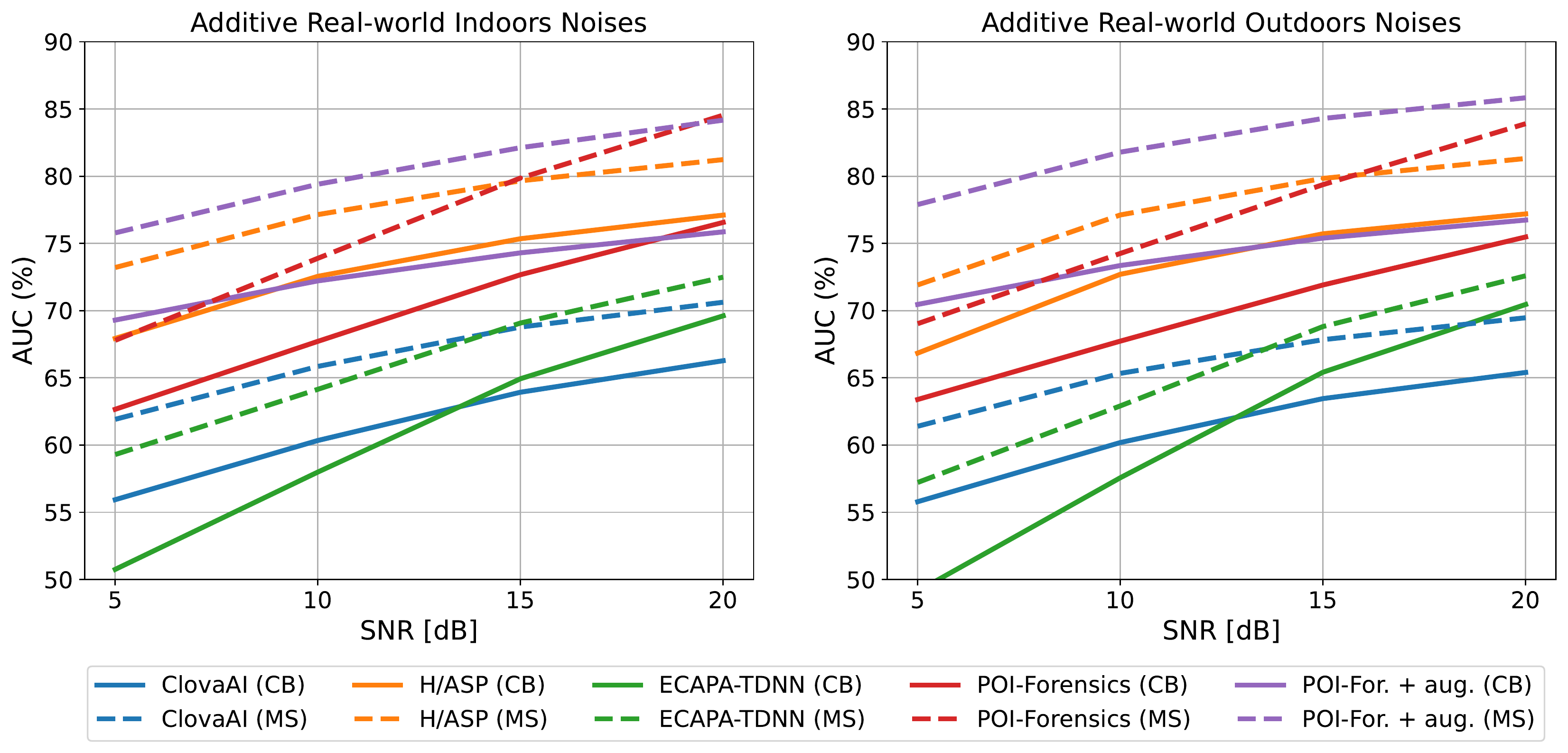}
    \caption{AUC performance on the IWA dataset in the presence of real noises as a function of the SNR for 
    the indoor (left) and outdoor (right) scenarios. 
    Solid and dashed lines are for the Centroid-Based and Maximum-Similarity testing strategies, respectively.}
    \label{fig:noise}
\end{figure*}

\subsection{Robustness analysis}

In figure \ref{fig:noise}, we show the AUC performance on the IWA dataset in the presence of background noise. 
We consider real-world noise using the environmental sounds of the ESC-50 dataset \cite{Piczak2015esc}.
In detail, we consider 
the sounds: breathing, footsteps, laughing, mouse-click, keyboard-type, and clock-tick in the indoor scenario and 
the sounds: engine, train, fireworks, rain, wind, and thunderstorm, in the outdoor scenario.
Of course, the performance worsens significantly when noise increases and, in these challenging condition,
the Maximum-Similarity strategy appears largely preferable.
The performance degradation is much reduced when augmentation is included in training. 
This appears clearly by comparing the red and purple curves,
corresponding to POI-Forensics and POI-Forensics with augmentation,
using background audios from the MUSAN dataset \cite{Snyder2015musan} with SNR going random from 3 to 25 dB.

\section{Conclusion}
Synthetic audio detection is drawing great attention, lately.
However, popular two-class methods, that perform very well on data generated by known methods, 
generalize poorly to data generated by new unseen methods, proving unfit to deal with real-world scenarios.
Here, we propose a new one-class strategy, based on speaker verification tools, which dispenses altogether with fake training audios and therefore provides both a good performance and an excellent generalization ability.

It may seem ironic that speaker verification is used for spoofing detection,
when spoofing detection methods were originally conceived to protect  speaker verification tools. 
However, these two approaches fit different scenarios, with different levels of prior information.
Our proposal represents just a new tool in the forensic analyst's hands to help counter malicious attacks.
However, preliminary results are already very promising, and we hope they will encourage research on similar solutions.

\section*{Acknowledgment}

We gratefully acknowledge the support of this research by the Defense Advanced Research Projects Agency (DARPA) under agreement number FA8750-20-2-1004. 

The U.S. Government is authorized to reproduce and distribute reprints for Governmental purposes notwithstanding any copyright notation thereon.
The views and conclusions contained herein are those of the authors and should not be interpreted as necessarily representing the official policies or endorsements, either expressed or implied, of DARPA or the U.S. Government. 

This work is also co-funded by the European Union under the Horizon Europe vera.ai project, Grant Agreement number 101070093,
and is supported by the PREMIER project, funded by the Italian Ministry of Education, University, and Research within the PRIN 2017 program. 

\newpage

\balance
\bibliographystyle{IEEEtran}
\bibliography{IEEEabrv,ref}

\end{document}